\documentclass[12pt]{article}
\input{psfig.sty}
\usepackage{graphicx}
\begin{document}

\title{Interpretation of experimental data
near $\lambda$-transition point in liquid helium}

\author{J. Kaupu\v{z}s 
\thanks{E--mail: \texttt{kaupuzs@latnet.lv}}
\hspace{1ex}\\
Institute of Mathematics and Computer Science \\ University of Latvia\\
29 Rainja Boulevard, LV--1459 Riga, Latvia}

\date{\today}

\maketitle

\begin{abstract}
The recently published experimental data for specific heat $C_p$
of liquid helium in zero gravity conditions very close to the
$\lambda$--transition have been discussed. We have shown
that these data allow different interpretations.
They can be well interpreted within the perturbative
RG approach and within our recently developed theory, as well.
Allowing the logarithmic correction, the corresponding fits lie 
almost on top of each other over the whole range of the reduced 
temperatures $t$ (for bin averaged data)
$6.3 \cdot 10^{-10} < t <  8.8 \cdot 10^{-3}$.
However, the plot of the effective exponent $\alpha_{\mbox{\scriptsize eff}}(t)$
suggests that the behaviour of $C_p$, probably, changes very close to $T_{\lambda}$.
To clarify this question, we need more accurate data for $t<10^{-7}$. 
In addition, we show that the experimental data
for superfluid fraction of liquid helium close to $T_{\lambda}$
within $t \in [3 \cdot 10^{-7};10^{-4}]$ can be better fit
by our exponents $\nu=9/13 \simeq 0.6923$, $\Delta=5/13 \simeq 0.3846$ 
than by the RG exponents $\nu \simeq 0.6705$ and $\Delta \simeq 0.5$.
The latter ones are preferable to fit the whole measured range  
$t \in [3 \cdot 10^{-7};10^{-2}]$ where, however, remarkable
systematic deviations appear.
Our estimated value $0.694 \pm 0.017$ of the asymptotic exponent $\nu$
well agrees with the theoretical prediction $\nu=9/13$.
\end{abstract}


\section{Introduction}

It is widely accepted to consider the measurements in liquid
helium near $\lambda$--transition point $T=T_{\lambda}$
as a crucial test of validity of the theoretical predictions for the critical
exponents, since these measurements are done with a high
degree of accuracy much closer to
the critical point than in any other experiments or numerical simulations.
In particular, it is believed that accurate experimental measurements of 
specific heat $C_p$ of liquid helium very close to the $\lambda$-transition
point  in zero gravity conditions (in space)~\cite{Lipa}
provide a convincing evidence of overall correctness
of the perturbative RG approach. 
The aim of our paper is to show that this conclusion is not
unambiguous, since these experimental data as well as those
of the superfluid fraction of liquid helium can be equally well 
or even better interpreted
by a completely different set of critical exponents provided
by our recently developed theory~\cite{K1}.

\section{Interpretation of the specific heat data}
\label{sec:Cp}

It has been found in~\cite{Lipa} that 
fits of experimental data for a wide range of reduced 
temperatures
$5 \cdot 10^{-10} \le t \le 10^{-2}$ below $T_{\lambda}$
by using two slightly different ansatz,
\begin{equation}
C_p = \frac{A^-}{\alpha} t^{-\alpha} \left( 1 + a_c^- t^{\Delta}
+ b_c^- t^{2 \Delta} \right) + B^-
\label{vinu}
\end{equation}
and
\begin{equation}
C_p = \frac{A^-}{\alpha} t^{-\alpha} \left( 1 + a_c^- t^{\Delta}
 \right) + b_c^- t + B^- \;,
\end{equation}
biased by the RG theoretical value of the correction--to--scaling
exponent $\Delta=0.529$, provide well consistent values of
the specific heat exponent $\alpha=-0.0127 \pm 0.0003$ in a
good agreement with the value $-0.01294 \pm 0.0006$ of the variational
perturbative theory~\cite{Kleinert} as well as in a worse, but still
acceptable, agreement with more recent estimates
$\alpha = -0.01126 \pm 0.0010$~\cite{Kl2} and
$\alpha = -0.0146 \pm 0.0008$~\cite{Camp}.
Apart from the exponent $\alpha$, some other quantities have been
determined and compared with
the RG values in~\cite{Lipa}. However, the agreement is not so good
to conclude that any theoretical approach, which does not agree with the
perturbative RG, is wrong. In particular, the experimental quantity
$P=(1-A^+/A^-)/\alpha$ is $4.154 \pm 0.022$, whereas the recent RG
calculation (Ref.~63 in~\cite{Lipa}) yields $P=4.433 \pm 0.077$.

We have found that the measured data of $C_p$ \cite{Lipa} can be well
reproduced also by an ansatz of the form
\begin{equation}
C_p = t^{-\alpha} (C + A \ln t) \left( 1 + a t^{\Delta} \right) + B \;,
\label{mans}
\end{equation}
with fixed exponents $\alpha=-1/13$ and $\Delta = 5/13$ proposed
in~\cite{K1,SPIE}. It is consistent with the idea that specific
heat can have a logarithmic correction, as discussed in~\cite{K1}.
The power--like singularity is recovered at $A=0$.
Note that in~\cite{K1} (cf.~Eq.~(60) there) an ordinary term 
$\sim t^{-\alpha}$ is related the behaviour of the correlation
function within the range of wave vectors 
$k \sim 1/\xi$, where $\xi \sim t^{-\nu}$ is the correlation length,
whereas the logarithmic term can appear due to the contribution of 
the region $k \gg 1/\xi$. In this aspect, the ratio $A/C$ in~(\ref{mans})
can be varied in a wide range of values. 

From the raw data of~\cite{Lipa} given in~\cite{dati} we have
produced the set of bin averaged data points by dividing each decade of the
reduced temperature $t = 1 - T/T_{\lambda}$ in 10 segments
of equal width when looking in the logarithmic scale.
One binned data point has been obtained by an averaging 
over $C_p$ and $t$ values within one segment, and only the data for 
the smallest $t$ values within a twice wider interval 
$4.7 \cdot 10^{-10}<t< 7.9 \cdot 10^{-10}$ have been merged together 
into one bin to reduce the statistical error. In our binning,
the averaged data points come as close to
$T_{\lambda}$ as $t \ge 6.3 \cdot 10^{-10}$,
whereas those given in~\cite{dati} extend only to $t \ge 7.94 \cdot 10^{-10}$.

\begin{figure}
\begin{center}
\includegraphics[scale=0.4]{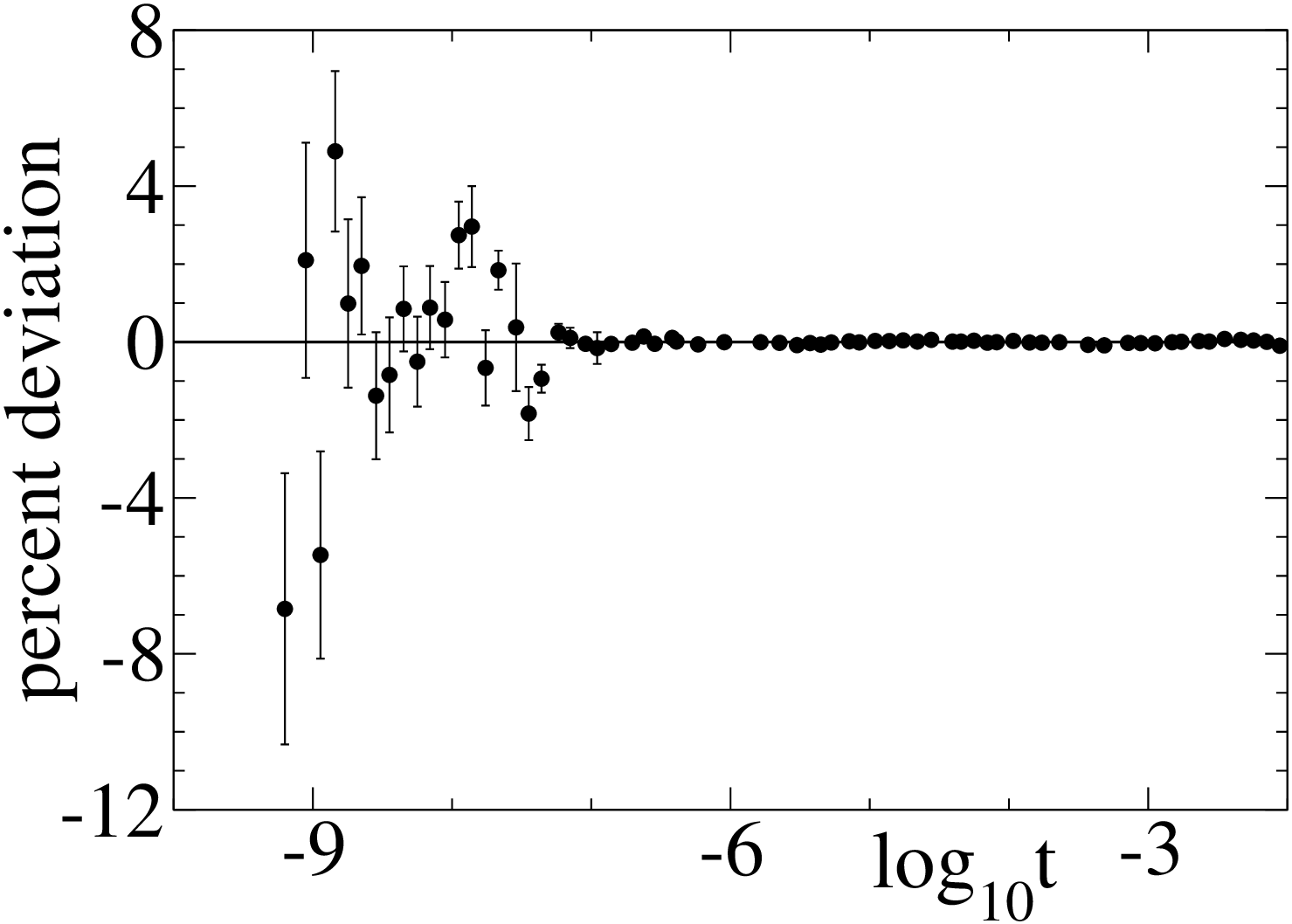}\\
\vspace{7mm}
\includegraphics[scale=0.4]{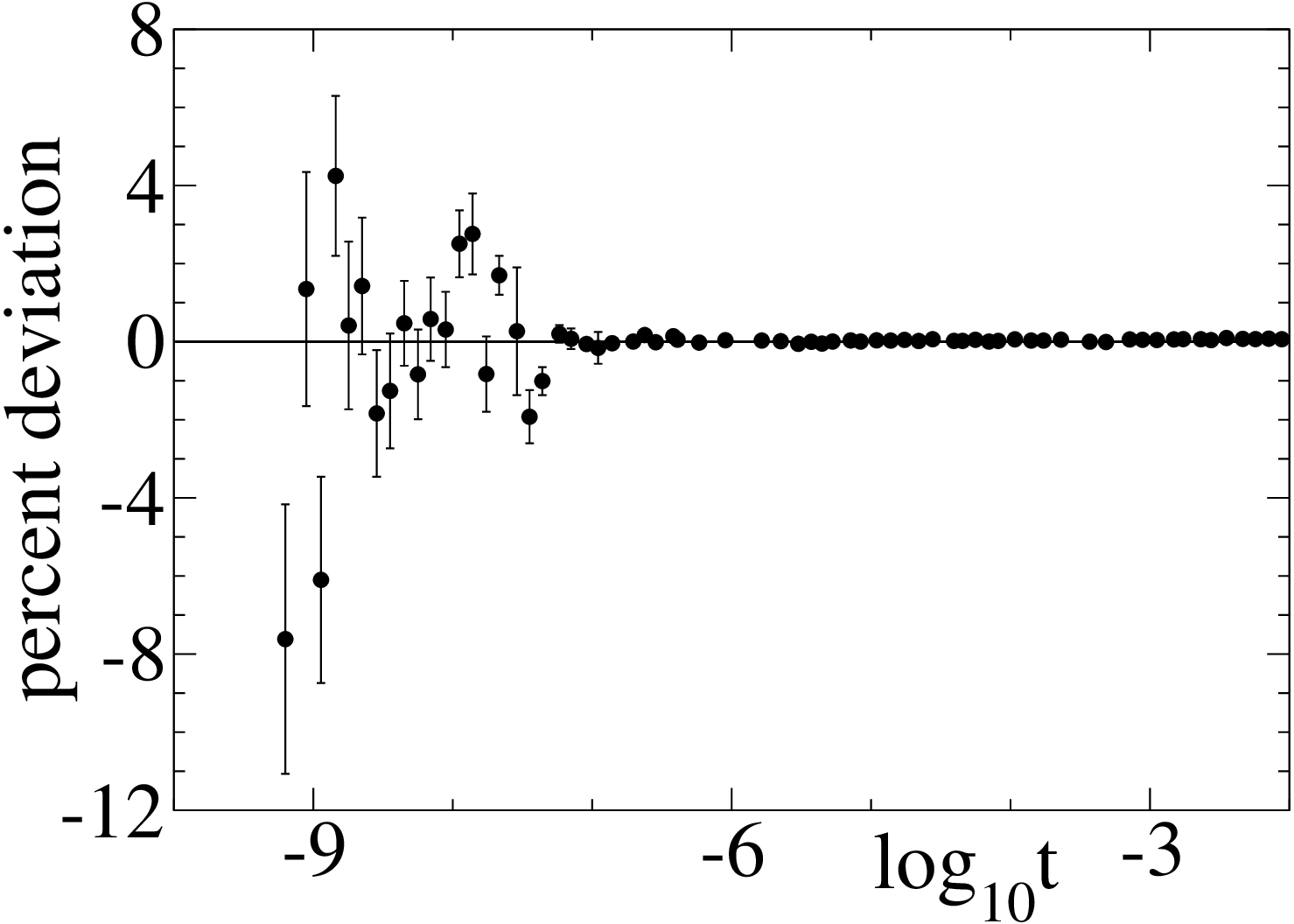}
\end{center}
\caption{Percent deviation of the fitted $C_p$ data points 
from the ansatz~(\ref{mans}) with fixed exponents
$\alpha=-1/13$ and $\Delta=5/13$ (top) and from the
ansatz~(\ref{vinu}) with the fit parameters found in~\cite{Lipa} (bottom).} 
\label{fits}
\end{figure}

The percent deviations from the least--squares fits to~(\ref{mans}) 
and~(\ref{vinu})
are shown in Fig.~\ref{fits}. The upper picture corresponds to~(\ref{mans})
with fixed exponents $\alpha = -1/13$ and $\Delta = 5/13$ and
coefficients $C=-167.536$, $A=11.6593$, $a=0.19788$, and $B=198.26$,
whereas the lower one represents the fit to~(\ref{vinu}) with
exponents $\alpha = -0.01264$ (fit parameter) and $\Delta=0.529$ (fixed) and 
coefficients listed in Tab.~II of~\cite{Lipa}. As in some fits made 
in~\cite{Lipa}, we have assigned the error bars $0.02\%$ to the bin averages
which originally had smaller errors.
In this way, we have reduced the impact of
these data points, located at relatively large values of the reduced 
temperature $t$, where the asymptotic ansatz~(\ref{mans}) is not very 
accurate. 
As we see from Fig.~\ref{fits}, both fits are almost identical in
the whole range of the reduced temperatures. The fit with our exponents
(top) is slightly worse at the largest $t$ values. It can be
well understood, since (to ensure the stability of the fit parameters)
we have neglected the subleading correction
of the kind $t^{2 \Delta}$ included in the other ansatz~(\ref{vinu}). 
Besides, our fit is even slightly better at the smallest $t$ values:
the mean percent deviation for 10 smallest $t$ values is $-0.425 \pm 0.690$
in our case of~(\ref{mans}) and $-0.975 \pm 0.686$ in the case
of~(\ref{vinu}). These deviations are reduced to $0.004 \pm 0.695$
and $-0.477 \pm 0.691$, respectively, when shifting the $T_{\lambda}-T$ values    
by $0.5$~nK within the experimental error bars~\cite{Lipa}.

Note that only the
possibility and not the necessity of the logarithmic correction follows from the
theory~\cite{K1}. However, the presence of the logarithmic correction for
specific heat, perhaps, is a quite general feature: the logarithmic singularity
(as a special case of the logarithmic correction when $\alpha=0$) 
of specific heat is a rigorously stated fact in 2D Ising model~\cite{Baxter}, and our
Monte Carlo simulation data for 3D Ising model~\cite{SPIE} also supports the
logarithmic singularity. Our analysis of the experimental data for the superfluid
fraction, made in Sec.~\ref{sec:supfl}, gives one more argument: it suggests
that the exponent $\nu$ is remarkably larger than $0.6705$. According to the
known scaling relation $\alpha + d \nu = 2$~\cite{Baxter}, the exponent
$\alpha$ then should be remarkably more negative than $-0.0115$. It well coincides
with the measured data~\cite{Lipa} in the whole range of the reduced temperatures only if the
pure power is perturbed by a logarithmic correction, as proposed by ansatz~(\ref{mans}).

Considering $\alpha$ as a fit parameter in~(\ref{mans}),
we obtain a value $\alpha = -0.0848 \pm 0.0039$, which is quite close
to our theoretical prediction $\alpha = -1/13 \simeq -0.0769$.
The small systematic deviation could be caused by the error
of the asymptotic ansatz~(\ref{mans}) at the largest $t$ values.
This problem cannot be reliably solved by
adding more correction terms or by narrowing the range of the fit,
since the minimum of the $\chi^2$ for such fits  
is very shallow, i.~e., the results become poorly defined.

Alternatively, we have fit the data within a moving window $t \in [t_i;100 t_i]$ 
of the reduced temperatures to the simplest possible ansatz
\begin{equation}
C_p = A \, t^{- \alpha} + B \;,
\label{eq:simplest}
\end{equation}
where $t_i$ corresponds to the $i$--th bin averaged data point.
It yields the effective exponent $\alpha_{\mbox{\scriptsize eff}}(t)$, where $t$ belongs
to the considered interval. For convenience, we have defined it as
$t = \sqrt{t_{min} t_{max}}$, where $t_{min} = t_i$ is the minimal and
$t_{max}$ is the maximal $t$ value in the interval. By this method,
the result converges to the true value of $\alpha$ at $t \to 0$
irrespective to the error of the asymptotic ansatz at finite $t$.
It works also when the logarithmic correction is present, only in this
case the convergence is very slow, like 
$\alpha - \alpha_{\mbox{\scriptsize eff}}(t) \sim 1/ \ln t$ at $t \to 0$. 
The only problem is that this method requires that the measurement errors
both for $C_p$ and $t$ remain sufficiently small when approaching $T_{\lambda}$.

\begin{figure}
\begin{center}
\includegraphics[scale=0.34]{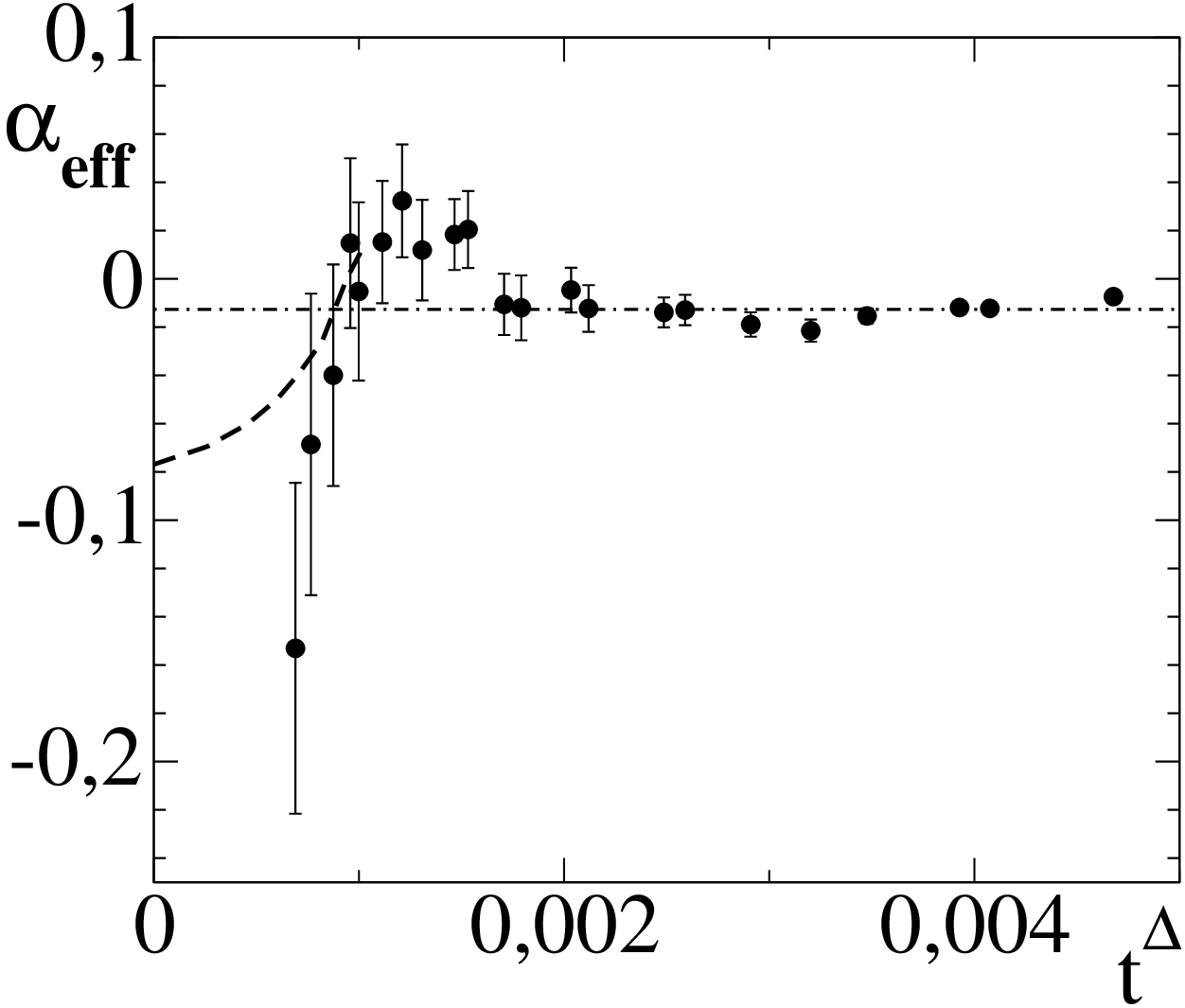}
\hspace{3mm}
\includegraphics[scale=0.34]{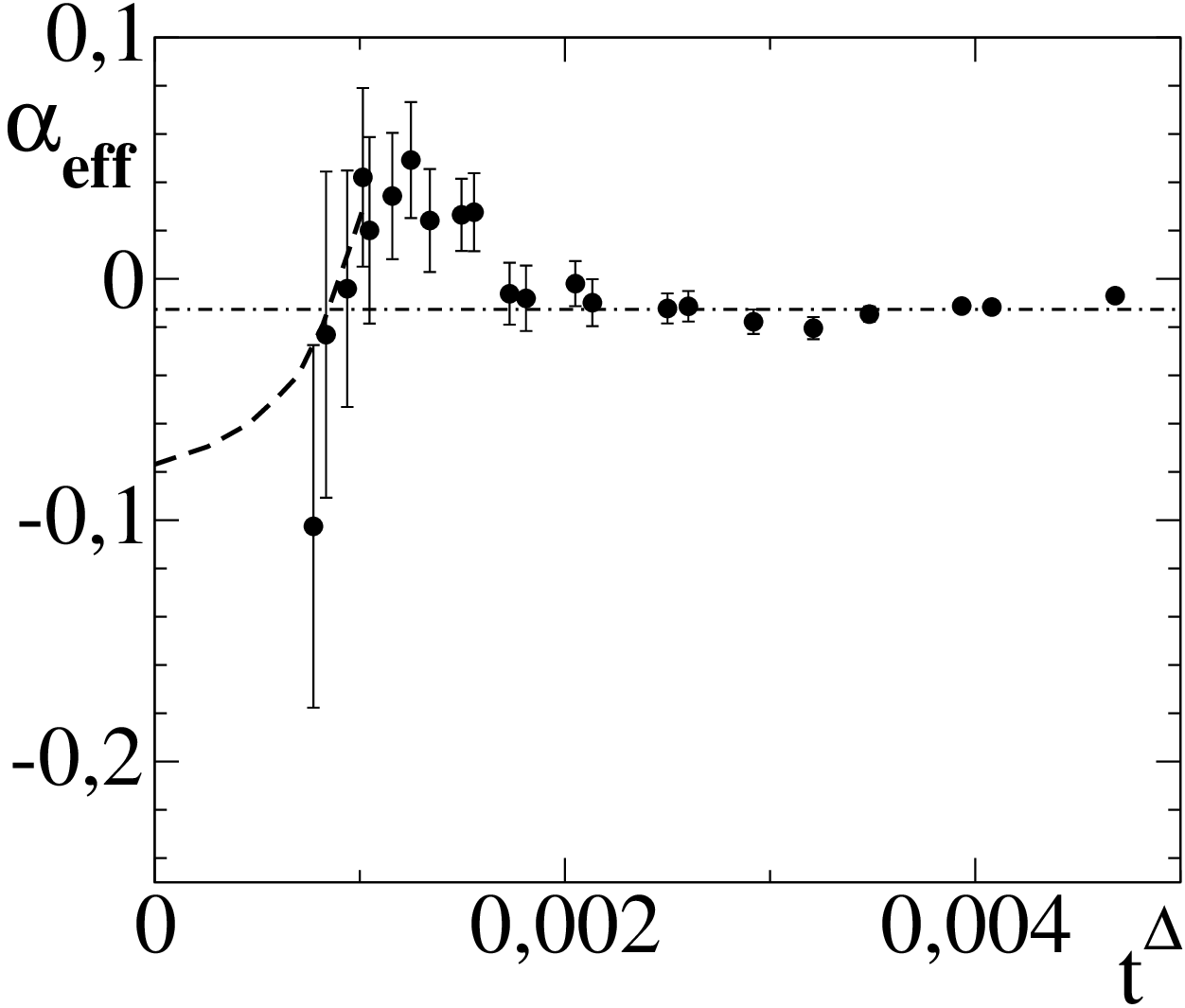}
\end{center}
\caption{The effective exponent 
$\alpha_{\mbox{\scriptsize eff}}$ vs $t^{\Delta}$
estimated by fitting  the bin averaged data within 
$[t_i;100t_i]$ to the ansatz~(\ref{eq:simplest}) with unbiased (left) and 
shifted according to 
$T_{\lambda} \to T_{\lambda} + 0.5~\mbox{nK}$ (right) values of the
reduced temperatures. The horizontal dot--dashed line indicates
the $\alpha$ value $-0.01264$ found in~\cite{Lipa}, whereas the dashed curve
shows schematically a behaviour expected from~\cite{K1}.}
\label{alfeff}
\end{figure}
We have plotted in Fig.~\ref{alfeff} the results for $\alpha_{\mbox{\scriptsize eff}}$ depending
on $t^{\Delta}$ (with our exponent $\Delta=5/13$) obtained by using the unbiased 
values of $t_i$ (left), as well as the shifted values 
$t_i' = t_i + 2.3 \cdot 10^{-10}$ (for the same sets of measured points)
corresponding to $T_{\lambda} \to T_{\lambda} + 0.5~\mbox{nK}$ (right). 
In spite of the large error bars, these plots show certain trend,
where the effective exponent $\alpha_{\mbox{\scriptsize eff}}(t)$ tends to decrease below
the (RG) value $-0.01264$ found in~\cite{Lipa}. Moreover, we have verified that 
the same trend is observed in both cases when only the odd and only the even
raw measurements (the original values listed in~\cite{dati}) are used. 
Hence, we cannot exclude any striking scenario. For instance, 
$\alpha_{\mbox{\scriptsize eff}}(t)$ could converge to our asymptotic value $\alpha = -1/13$, 
as indicated by dashed lines, particularly, if we allow a small shift
in $T_{\lambda}-T$ values (right picture) within the experimental error bars.
However, such a behaviour would mean that the logarithmic correction
is absent, since the convergence is rather fast. It would imply also that
the estimation of $\alpha$ from the fit over the whole measured range is not valid:
formally, such a fit looks good, but it effectively ignores the systematic deviations 
at the smallest $t$ values where the error bars are larger. 

Due to the experimental errors,
the results of our analysis of the effective exponent are not conclusive, 
they only point to a possible scenario. From an intuitive point of view,  
it does not seem plausible that such a remarkable change in the behaviour of
the system could take place at so small reduced temteratures ($t<10^{-7}$). 
However, since the deviations from the asymptotic scaling law
are caused by the corrections to scaling, an essential parameter is 
$t^{\Delta}$ rather than $t$, and the values of $t^{\Delta}$ in Fig.~\ref{alfeff} are
not so extremely small. Besides, the critical region, where an asymptotic 
ansatz is valid, can be as narrow as $t^{\Delta} \sim 0.001$ even in a simple 
mean field model. An example is given in~\cite{PhysRep} (p.~75).
From this point of view, it is possible that the deviations in Fig.~\ref{alfeff}
represent a real physical effect and not an artifact.
On the other hand, random deviations in Fig.~\ref{fits} too often are as large as
$2$ standard deviations or even larger, therefore the unusual behaviour
of $\alpha_{\mbox{\scriptsize eff}}$ in Fig.~\ref{alfeff} can be ascribed also as an artifact.

\section{Interpretation of the experimental data for\\ the superfluid fraction}
\label{sec:supfl}

Here we discuss the experimental data for the superfluid fraction $\rho$
in liquid $He$. It decreases asymptotically 
(at $t \to 0$) as $\rho \sim t^{\zeta}$.
It is believed (see~\cite{GA} and references therein) that
the exponent $\zeta$ is equal to the correlation length exponent $\nu$ 
for the 3D $XY$ model. In~\cite{K1}, the superfluid fraction of $^4 He$
measured in~\cite{GA} has been discussed with an aim to
compare the experimentally observed behaviour at the temperatures closest to
$T_{\lambda}$ with our theoretical prediction $\nu = 9/13$~\cite{K1}.

The data listed in~\cite{GA1} allow a more precise comparison.
For this purpose, first we have fit these data to the asymptotic ansatz
\begin{equation}
\rho = A \, t^{\nu} \left( 1+ a_1 t^{\Delta} + a_2 t^{2 \Delta} \right)
\label{ansatz} 
\end{equation} 
including two corrections to scaling. Similar fits over the
whole measured range $t \in [3 \cdot 10^{-7}; 10^{-2}]$ have been
considered in~\cite{GA,GA1}. Note that at $\Delta = 0.5$, used 
in~\cite{GA,GA1}, the second order correction reduces to the analytical
one, and~(\ref{ansatz}) differs
from the ansatz of~\cite{GA,GA1} only by a remainder term of higher order. 
The overall fits discussed in~\cite{GA,GA1} yield $\nu \simeq 0.6705$
in agreement with the RG prediction and in disagreement with
our value $\nu=9/13$. However, these fits look really
good only within $t \in [10^{-5};10^{-2}]$, whereas remarkable
systematic deviations appear at smaller $t$ values. This phenomenon was 
discussed in~\cite{GA1} and no reasonable explanation was found.
In particular, the effect of gravity is negligible in these experiments~\cite{GA1}
and the $\pm 20$~nK uncertainty in $T_{\lambda}-T$ also does not explain
these systematic deviations.

\begin{figure}
\begin{center}
\includegraphics[scale=0.35]{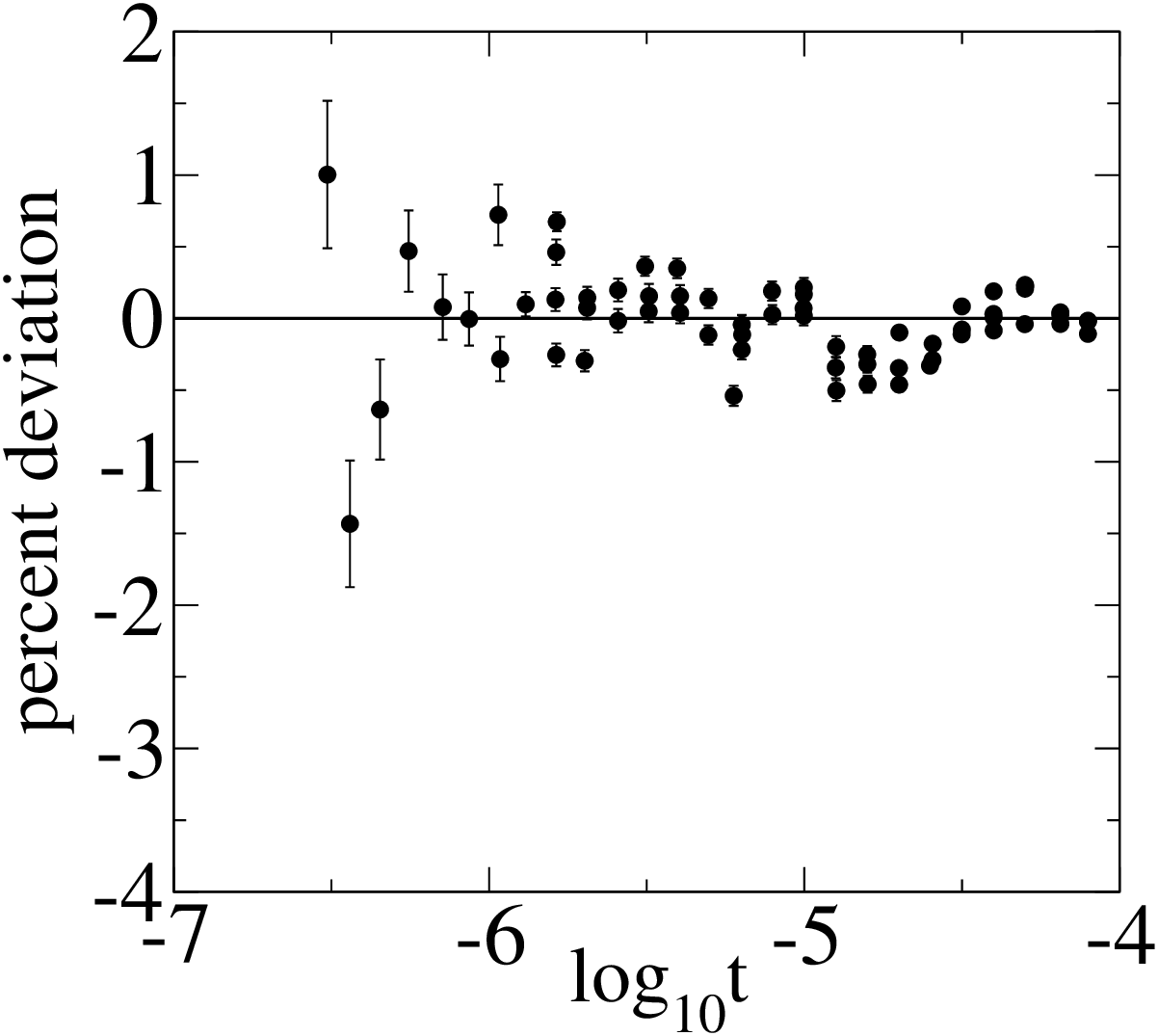}
\hspace{5mm}
\includegraphics[scale=0.35]{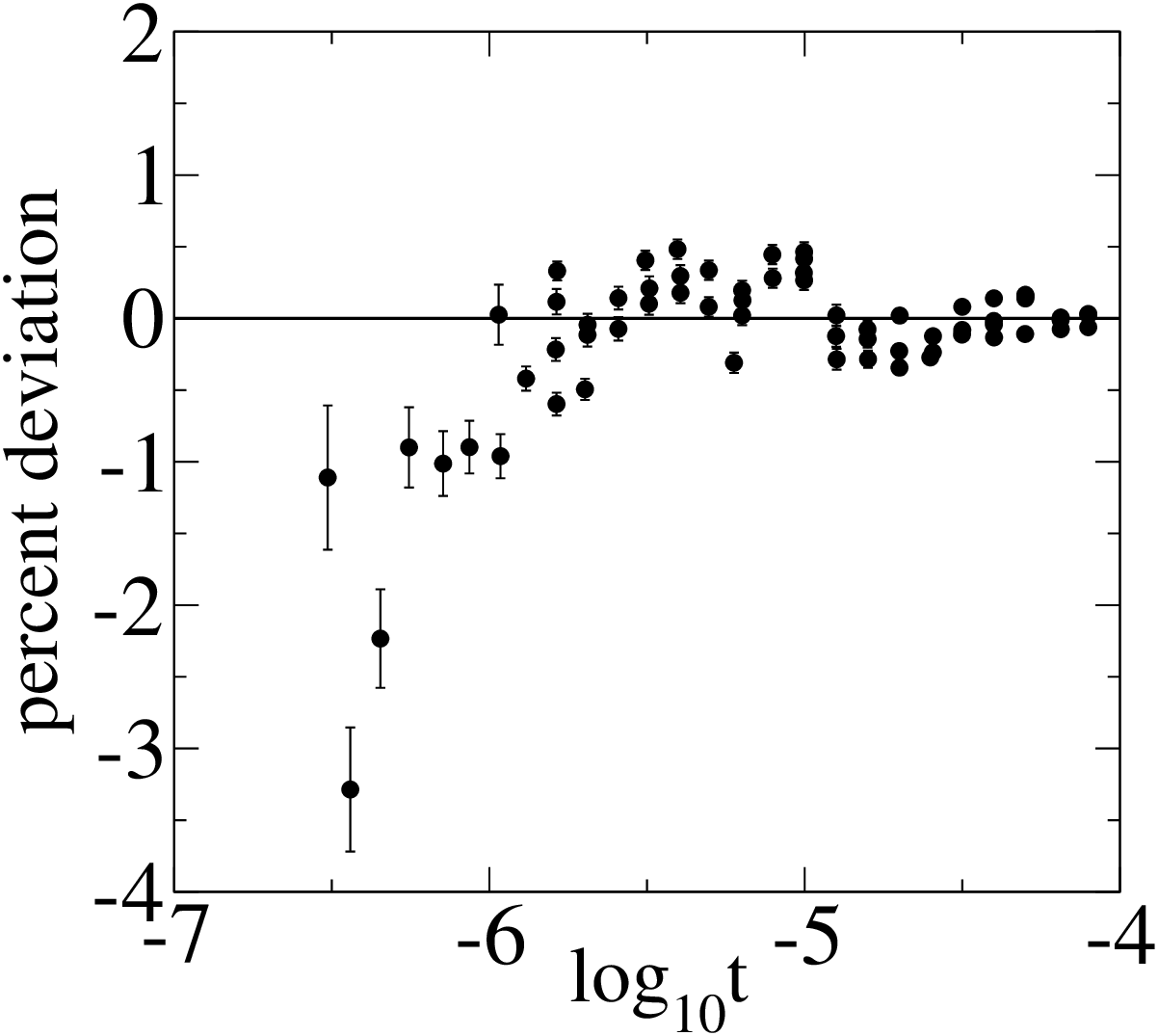}
\end{center}
\caption{Percent deviation of the experimental $\rho$ (superfluid fraction)
data points from the least squares fit to ansatz~(\ref{ansatz})
within $t \in [3 \cdot 10^{-7};10^{-4}]$
with our exponents $\nu=9/13$, $\Delta=5/13$ (left)
and with the RG exponents $\nu=0.6705$, $\Delta=0.5$ (right).}
\label{rhofits}
\end{figure}

Our theory~\cite{K1} provides an explanation. First, the data cannot be
well fit within the whole measured range $t \in [3 \cdot 10^{-7}; 10^{-2}]$ simply 
because the critical region, where the asymptotic expansion
in $t$ powers is valid, is much narrower than $10^{-2}$. Second, 
the data can be well fit to~(\ref{ansatz}) with our exponents
$\nu = 9/13$ and $\Delta=5/13$ within a reduced range $t \in [3 \cdot 10^{-7};10^{-4}]$,
which means that the measured data for $t<10^{-5}$ are not anomalous, 
but the region of validity of~(\ref{ansatz}) is as narrow as $10^{-4}$. 
The percent deviations from the least--squares fits within $t \in [3 \cdot 10^{-7};10^{-4}]$
with our (left) and RG (right) exponents are shown in Fig.~\ref{rhofits}.
As we see, in our case there are no essential systematic deviations, whereas in the
RG case they are observed like in the case of
the fit over the whole measured range~\cite{GA,GA1}.

\begin{figure}
\begin{center}
\includegraphics[scale=0.48]{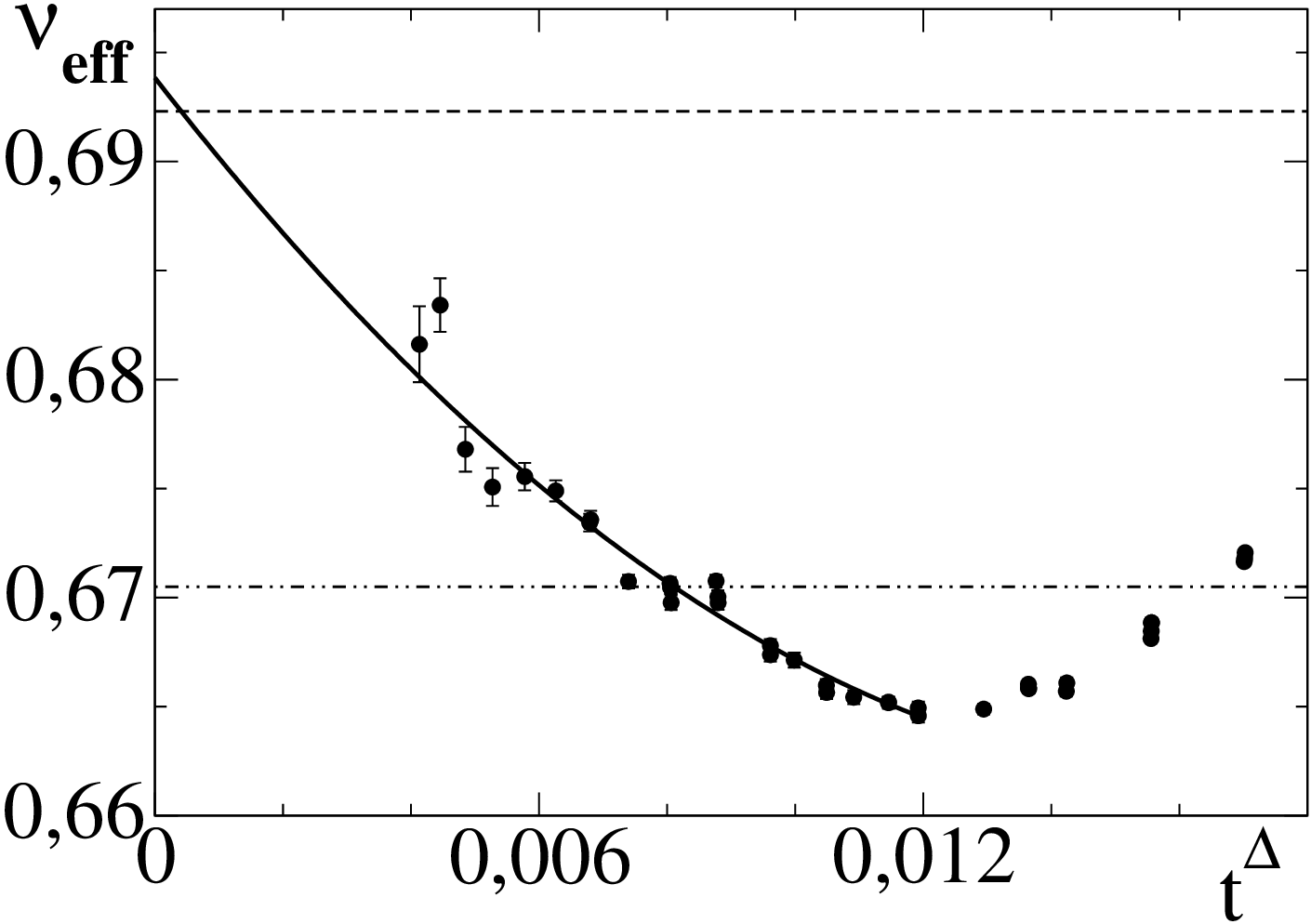}
\end{center}
\caption{The effective exponent 
$\nu_{\mbox{\scriptsize eff}}$ vs $t^{\Delta}$ determined
from the local slopes of the $\ln \rho$ vs $\ln t$ plot. The lower dot-dot-dashed
line indicates the (RG) value $0.6705$ obtained in~\cite{GA,GA1}, whereas
the upper dashe line shows our theoretical value $\nu=9/13$. The solid curve represents
the least--squares fit to a parabola.}
\label{nueff}
\end{figure}

Similarly as in Sec.~\ref{sec:Cp}, we have evaluated also the effective
exponent $\nu_{\mbox{\scriptsize eff}}(t)$ as the local slope of the $\ln \rho$ vs $\ln t$  plot within 
$[t_i;5 t_i]$, where $t_i$ is the reduced temperature of the $i$--th measurement and $t$ is the
middle point of the fitted interval in the logarithmic scale.
The results depending on $t^{\Delta}$ (with our value $\Delta=5/13$) 
are shown in Fig.~\ref{nueff}. Evidently, the effective exponent
tends to deviate above the value $0.6705$ (dot-dot-dashed line) 
obtained in~\cite{GA,GA1}. 
On the other hand, the fit of this plot to a parabola (solid line)
gives the asymptotic estimate $\nu = 0.694 \pm 0.004$ in excellent
agreement with our theoretical value $9/13 \simeq 0.6923$. However, taking into account
the $\pm 20$~nK uncertainty in the $T_{\lambda}$ value~\cite{GA}, the error bars become larger,
i.~e., $\nu = 0.694 \pm 0.017$.

\section{Conclusions}

Although the opinion dominates in publications that the perturbative RG theory 
is strongly confirmed by very accurate measurements of the specific heat and the superfluid 
fraction in liquid helium near the $\lambda$--transition point, our current analysis shows
that these experimental data can be well understood and interpreted also within
our recently developed theory~\cite{K1}. 

In summary we conclude the following:
\begin{enumerate}
\item 
The critical exponents of the perturbative RG theory look preferable from a point of view that 
all measured data points must be necessarily fit on one curve and no logarithmic
corrections are normally expected. However, if we allow the logarithmic correction
to specific heat $C_p$, our theory also provides a good fit of $C_p$ data for the whole 
measured range. Our fit then is slightly worse at the largest reduced temperatures 
$t \sim 10^{-2}$ and better at the smallest ones $t \sim 10^{-9}$. 
As discussed in Sec.~\ref{sec:Cp}, the existence of the logarithmic correction is partly 
supported by our estimation of the exponent $\nu$ (Sec.~\ref{sec:supfl}), 
as well as by some general argument.

\item
The analysis of the effective exponent $\alpha_{\mbox{\scriptsize eff}}$
indicates that the behaviour of $C_p$ could be remarkably changed
very close to $T_{\lambda}$, in such a way that the true asymptotic singularity is
power--like (without the logarithmic correction) with the exponent which more probably 
is closer to our value $\alpha =-1/13$ than to the (RG) value $-0.01264$ obtained in~\cite{Lipa}.
However, this effect can be ascribed also as an artifact caused by the measurement errors.
Further improvement of the experimental accuracy for $t<10^{-7}$ would be very helpful 
to clarify this question.

\item
As compared to the RG exponents, our critical exponents are better consistent with the 
closest to $T_{\lambda}$ data ($t \in [3 \cdot 10^{-7}; 10^{-4}]$) 
for the superfluid fraction (cf.~Figs.~\ref{rhofits} and~\ref{nueff}).
A self consistent estimation in this case yields $\nu = 0.694 \pm 0.017$ in agreement
with our theoretical prediction $\nu = 9/13 \simeq 0.6923$. Since the fit over the whole measured range
($t \in [3 \cdot 10^{-7}; 10^{-2}]$) in no case is really good, we argue that our way of 
estimation is preferable.
\end{enumerate}

\section*{Acknowledgments}
This work partly has been done during my stay in 2005 at the Institute of Physics of 
Rostock University, Germany.

\end{document}